\documentclass[12pt]{emulateapj}
\usepackage{lscape}

\submitted{Accepted by Astrophysical Journal, 2005 September 2}
\newcommand{\etal}{{\it et al.}}

\shorttitle{H$\alpha$ Atlas of Isolated Spirals}

\begin{document}

\title
{An Atlas of H$\alpha$ and R Images and Radial Profiles of 29 Bright Isolated Spiral Galaxies}

\author{Rebecca A. Koopmann}
\affil{Union College}
\affil{Department of Physics and Astronomy, Schenectady, NY 12308}
\email{koopmanr@union.edu}

\author{Jeffrey D. P. Kenney}
\affil{Astronomy Department}
\affil{Yale University, P.O. Box 208101, New Haven, CT 06520-8101}
\email{kenney@astro.yale.edu}

\begin{abstract}
Narrow-band H$\alpha$+[NII] and broadband R images and surface photometry
are presented for a sample of 29 bright (M$_B < $ -18) 
isolated S0-Scd galaxies within a distance of 48 Mpc. 
These galaxies are among the most isolated nearby spiral galaxies of 
their Hubble classifications as
determined from the Nearby Galaxies Catalog (Tully 1987a). 
\end{abstract}


\section{Introduction}

How does the environment of a galaxy influence its evolution? 
Finding an answer to this question requires extensive 
comparisons between galaxies in different environments,
including those in the most isolated environments. 

Many studies have sought to compare the global star formation 
properties of galaxies in field and cluster environments (see, for example,
references in Koopmann \& Kenney 2004a and a review by Poggianti 2005). 
The selection of isolated comparison samples in these studies is
as important as the selection
of the primary objects of study. In the case of a study on environmental
effects, the comparison sample should ideally be made up of galaxies
completely isolated from their surroundings. However, the distribution 
of galaxies in space appears to be strongly clustered, with few, 
if any, objects
outside gravitationally associated structures. A truly isolated galaxy
may not exist (e.g., Vettolani \etal ~1986, Tully 1987b). All is not lost,
however. The main requirement for a galaxy to serve as a comparison object in
a study of the environmental effects on bright cluster galaxies is that it 
have a low probability of interaction with another galaxy of similar mass
or with a group environment over a Hubble time. Such galaxies are not
typical (since 69\% of galaxies are located in groups; Tully 1987b), but
they do exist.

We undertook to observe a sample of isolated galaxies as part of 
a study to compare the star formation properties of nearly 100 bright isolated
and Virgo Cluster spiral galaxies via broadband R and H$\alpha$ images. 
We sought the most isolated nearby galaxies for our comparison sample,
using the environmental information in the Nearby Galaxies Catalog 
(Tully 1987a, hereafter NBG). 
Because we wished to compare Virgo and isolated spirals as a function of
Hubble type classification, we selected our sample based on
Hubble type as well as local galaxy density, 
observing roughly equal numbers of different spiral types,
with luminosities, distances, and inclinations
as similar as possible to those of Virgo spirals.
We use the term `isolated' for this sample rather than the
more general term `field'. The selection of field samples is
generally not subject to the stringent requirements outlined here, and
many field galaxy samples contain group members.

Since our primary motivation was to find an isolated sample suitable for
comparison to Virgo Cluster galaxies,
our selection process was somewhat different from that of other
studies which have sought to identify and characterize the most
isolated galaxies. Studies such 
as Karachentseva (1973) and Pisano, Wilcotts, \& Liu (2002) apply
more stringent density limitations, but their
samples contain relatively few early-type spiral galaxies and are on 
average further away. We note that much 
larger samples of isolated galaxies are being compiled
through studies such as Allam \etal ~(2005), who
identify a sample of 2980 isolated galaxies in the Sloan Digital Sky 
Survey (SDSS; York et al. 2000) Data Release 1. The galaxies identified
in this survey are on average at much greater distance 
than Virgo, but the sample contains a significant number of
early-type galaxies.

A dilemma in selecting isolated galaxies for comparison was deciding
whether or not to select only galaxies which had previous measurements at other
wavelengths. For example, it would be useful to have a 
comparison sample of galaxies also mapped in HI and CO. However, we decided
that such a selection criterion might bias the sample, since galaxies
with mapped HI and CO were often selected because of their peculiarities.
Our final sample contains rather obscure galaxies, but
we hope that this will be changed in the future by further observations
of these galaxies. 

In this paper, we present images and radial profiles in H$\alpha$ and
R for 29 isolated spiral galaxies.  There are several other papers in
this series.  R and H$\alpha$ images and selection details of the
Virgo Cluster sample are presented in Koopmann, Kenney, \& Young (2001). 
Comparisons between the R and H$\alpha$ radial profiles of Virgo
Cluster and isolated spirals are presented in Koopmann \& Kenney
(1998; 2004a, 2004b).

This paper is organized as follows.
The selection of the isolated galaxies is described in Section~\ref{select}.
Section~\ref{observe} describes the observations and
reduction procedures. Section~\ref{atlas} presents the atlas of
isolated galaxies, including images and surface photometry.

\begin{deluxetable*}{lll|ll}
\tabletypesize{\scriptsize}
\tablecaption{Completeness of Observed Isolated Sample}
\tablewidth{0pt}
\tablehead{
 &\multicolumn{2}{c}{\Large{RSA}}&
\multicolumn{2}{c}{\Large{RC3}}\\
\tableline
\colhead{Hubble Type}
&\colhead{M$_B$ $\le$ -19}
&\colhead{-19 $<$ M$_B$ $\le$ -18}
&\colhead{M$_B$ $\le$ -19}
&\colhead{-19 $<$ M$_B$ $\le$ -18}}
\startdata
S0-S0/a&\ 50\% (4/8)&  100\% (1/1)&     36\% (4/11)&  50\% (1/2)\\
Sa-Sab&\  44\% (4/9)& \ \ - \ \ (0/0)&  60\% (9/15)& \ 0\% (0/1)\\
Sb-Sbc&   27\% (6/22)& 33\% (1/3) &     19\% (5/26)&  13\% (1/8)\\
Sc-Scd&   21\% (7/34)& 22\% (2/9)&      15\% (6/41) & 18\% (2/11)\\ 
\tableline
Total (all)& \        29\% (21/73)&  11\% (4/13)& 26\% (24/93)&  18\% (4/22)\\ 
Total (Sa-Scd)& \ 26\% (17/65)&  18\% (3/12)& 24\% (20/82)&  15\% (3/20)\\
\tableline
Total (all mags) &     29\% (25/86)\tablenotemark{a} & - &   24\% (28/115) \tablenotemark{b}&- \\
Total (Sa-Scd) & \ 26\% (20/77) & - &   23\% (23/102)& -
\enddata
\label{tabisocomp}
\tablecomments{Completeness of the observed
isolated sample given as percentages and
ratios of observed to total. The following selection criteria were used to
extract candidate galaxies from the NBG catalog: 
no group members, v$_{he} <$ 2500 km s$^{-1}$, distance 
$<$ 33 Mpc, NBG local density $<$ 0.3 gal Mpc$^{-3}$, inclination 
$<$ 75$^{\circ}$. Completeness is given as a function of both RSA and
RC3 morphological types. Absolute magnitudes are derived from distances
given by NBG.}
\tablenotetext{a}{The total number is 25 rather than 29 because four 
galaxies were not listed in the RSA catalog.}
\tablenotetext{b}{The total number is 28 rather than 29 because one galaxy
classified as Sc by RSA was classified as Sm by RC3.}
\end{deluxetable*}

\section{Sample Selection}
~\label{select}
An isolated comparison sample of galaxies should resemble the cluster sample as
much as possible in distance and absolute magnitude, but be located in the 
lowest density regions possible. The
selection of isolated galaxies was based on the environmental information 
contained in the NBG, supplemented by group listings from the work of
Gourgoulhon \etal ~(1992).

The NBG lists 2367 galaxies with recessional velocities $<$ 3000 km
s$^{-1}$. The catalog is essentially complete for galaxies with
recessional velocities $<$ 1500 km s$^{-1}$.  The galaxies in the
catalog are assigned memberships within hierarchical structures
ranging from large-scale `clouds' to associations and
groups. Associated galaxies must meet a critical density requirement
to be considered a bound group with a crossing time shorter than a
Hubble time. Unbound `associations' of galaxies meet a density
requirement one order of magnitude lower than the group requirement.
Our first step towards finding the most isolated galaxies was to
eliminate all group members, which amounts to approximately 69\% of
the NBG (Tully 1987b).  The remaining galaxies were either in
associations (20\%) or `at-large' within a cloud, i.e., apparently
unassociated with other cloud members (10\%).  Selection of our sample
from these galaxies was based on the Hubble type, local density,
distances, recessional velocities, inclinations, and luminosities
provided in the catalog (described in detail by Tully 1988).

We selected only spiral galaxies with Hubble types S0-Scd, and sought
to observe sufficient galaxies of each spiral type to provide a
comparison sample for Virgo Cluster galaxies. Thus the observed
galaxies do not represent the actual number distribution in terms of
Hubble type.  Only galaxies with inclinations of $< 75^{\circ}$ were
selected.  The luminosity cutoff was set to the same as that of the
Virgo Cluster sample: M$_B < $ -18.  The local density, distance, and
recessional velocity limits were determined by experimentation to
obtain sufficient numbers of early-type spirals, which are less common
in field environments.  To select galaxies with distances comparable
to Virgo, a cutoff in velocity of 2500 km s$^{-1}$ and a cutoff in
distance of 33 Mpc (as listed by NBG) were applied.  66\%
($\frac{19}{29}$) of the galaxies have v$_{\rm los}$ $<$ 1500 km
s$^{-1}$ and 86\% ($\frac{25}{29}$) $<$ 2000 km s$^{-1}$.  Galaxies
with v$_{\rm los}$ $>$ 1500 km s$^{-1}$ are mostly early-type spirals.
The distances of the galaxies, originally extracted from the NBG, were
reevaluated using the heliocentric velocity and the multiattractor
model for the velocity fields within the Local Supercluster (Tonry
\etal ~2000; Masters, K. L., private communication), with an assumed
Hubble constant of 70 km s$^{-1}$ Mpc$^{-1}$. 37\% of the galaxies are
within 20 Mpc, while 62\% are within 25 Mpc.

Our absolute density cutoff was set at 0.3 gal Mpc$^{-3}$.  The local
density of a particular galaxy is based on the contributions of all
galaxies with absolute magnitudes brighter than -16, including the
galaxy itself.  Local densities range from 0.06 gal Mpc$^{-3}$ for an
isolated galaxy to 4.2 gal Mpc$^{-3}$ in the central regions of the
Virgo Cluster.  For comparison, the density of the Local Group is 0.52
gal Mpc$^{-3}$. The local density is at least 0.1 gal Mpc$^{-3}$ for
90\% of the galaxies of all classes, meaning that almost all galaxies
have companions within $\sim$ 1 Mpc. (It should be emphasized that the
Nearby Galaxies Catalog is not able to identify galaxies with small
companions, i.e., companions with M$_B$ $>$ -16.  However, our
objective is to select galaxies which are the least likely to have
experienced similar mass interactions.)  In the observed sample, 69\%
of the 29 galaxies selected from the Nearby Galaxies Catalog have
local densities $\le$ 0.2 gal Mpc$^{-3}$ and 34\% $\le$ 0.1 gal
Mpc$^{-3}$.  67\% of the sample galaxies appear to belong to a
specific cloud, but are 'at-large' in the cloud, and 33\% belong to
associations within clouds.

A total of 115 isolated galaxies matching our selection criteria were
found in the NBG.  We provide in Table~\ref{tabisocomp} a summary of
the percentages of galaxies observed, according to the classifications
given by the two main catalogs: (i) \it Revised Shapley Ames Catalog
\rm (Sandage \& Tammann 1987; hereafter RSA) or the \it Carnegie Atlas
\rm (Sandage \& Bedke 1994; hereafter CA) and (ii) \it Third Reference
Catalog \rm (deVaucouleurs \etal ~1991; hereafter RC3). (Note that
only 86/115 of the NBG galaxies appeared in the RSA/CA.)

Additional targets were identified from a list of 86 isolated galaxies
derived from analyses described in Gourgoulhon \etal ~(1992) and
Fouqu\'{e} \etal ~(1992) and kindly provided to us by
P. Fouqu\'{e}. The list is derived from a sample of 4143 galaxies with
blue isophotal diameters, D$_{25}$, larger than 100$^{\prime\prime}$
and recessional velocities $<$ 6000 km s$^{-1}$, with completeness of
84\%.  Group and association members were identified in a manner
similiar to the NBG, but with a somewhat different approach to
determination of critical density and corrections for faint
galaxies. The final percentage of group members ($\sim$ 65\%) is
similar to the NBG.  Six of the observed galaxies appeared in both
lists. One galaxy which appeared only in the Fouqu\'{e} list
(NGC~2640) was also observed, but not included in the final analysis
due to reduction problems.

The final selection of galaxies was also dependent on the time of year
allocated for observing. 

The properties of the galaxies presented are listed in
Table~\ref{tabsample}.  The columns of Table~\ref{tabsample} contain
the following information: (1) Name of galaxy, (2) and (3) galaxy
coordinates, (4) RSA/CA Hubble type, (5) RC3 Hubble type (6) the
total, face-on blue magnitude from deVaucouleurs \etal ~(1991), (7)
the heliocentric radial velocity, (8) the distance of the galaxy in
Mpc, based on the heliocentric velocity and the multiattractor model
for the velocity fields within the Local Supercluster (Tonry \etal
~2000; Masters, K. L., private communication), with an assumed Hubble
constant of 70 km s$^{-1}$ Mpc$^{-1}$, (9) A$_R$, scaled from A$_B$ values
listed in deVaucouleurs \etal ~(1991) using values given in 
Riefe \& Lebofsky (1985), 
(10) the HI deficiency parameter, as defined by Solanes
\etal ~(1996; see also Giovanelli \& Haynes 1983), (11) the local
galaxy density in units of galaxies Mpc$^{-3}$, given in the NBG, and
(12) the affiliation of a galaxy with respect to associations, as
defined by the NBG: 'at-large' means the galaxy is unassociated with
other galaxies within the large scale cloud, 'association' means that
the galaxy is associated with other galaxies within the cloud.  The HI
fluxes used to extract the HI deficiencies in column (10) were
extracted from the Cornell University Extragalactic Group private
digital HI archive, and are corrected for effects such as beam
dilution (Springob \etal ~2005; see also Haynes \& Giovanelli 1984),
while optical diameters were obtained from the Arecibo General
Catalog, a private database maintained at Cornell University by Martha
Haynes and Riccardo Giovanelli.

Both the RC3 and RSA classes are
listed in Table~\ref{tabsample}. 
(Note that the agreement between classifiers is far better for this
isolated sample of galaxies than for galaxies listed in the Virgo Cluster 
sample of Koopmann \etal ~2001). 

\section{Observations and Reductions}
\label{observe}
The observations of the isolated galaxies were made on the KPNO 0.9-m
and the CTIO 0.9-m and 1.5-m telescopes on photometric nights between
August 1993 and February 1996. Exposures ranged from 3-15 min in a
standard Harris (Kron-Cousins) R filter, and 1-2 hrs in an H$\alpha$
filter of width 60-80 \AA, centered either near 6573 \AA \ or near
6600 \AA \ (for galaxies with v $>$ 1000 km/s).  An observing log is
presented in Table~\ref{tabobslog}. Galaxies are listed in order by
right ascension, with two to three lines describing the R and
H$\alpha$ observations. The columns list the following information:
(1) name of the galaxy, (2) the date of the observation, (3) the
telescope and chip in column 3 (where chip characteristics are given
in Table~\ref{chip}), (4) filter (where filter codes and
characteristics are given in Table~\ref{filter}), (5) exposure time in
seconds, (6) airmass of the observation, (7) the full width half
maximum (FWHM) in arcseconds, (8) the sky background sigma, (9) and
the estimated uncertainty in the sky level. The units of columns (8)
and (9) are erg cm$^{-2}$s$^{-1}$arcsec$^{-2}$.  The values for the R
observations may be converted to magnitudes using a zeropoint of
13.945.

The absolute flux calibration was based on observations of the
spectrophotometric standard stars from the lists of Massey \etal ~(1988)
and Hamuy \etal ~(1992). Landolt (1992) 
standards were observed to derive the extinction coefficient and/or
monitor the photometricity, and in some cases, to calibrate the observations.

Reduction procedures were identical to those described in Koopmann
\etal ~(2001).  Multiple exposures were registered and combined.
Images were flux calibrated to the Massey \etal ~(1988) and Hamuy
\etal ~(1992) spectrophotometric standards.  The sky background was
subtracted from the R and H$\alpha$ images before continuum
subtraction. Uncertainties in the sky levels were typically 0.5-1\%
for R and 1-2\% for H$\alpha$, as listed in Column (9) of
Table~\ref{tabobslog}. The continuum light was subtracted from the
H$\alpha$ image using a scaled R image. A full discussion of the
uncertainties in the continuum subtraction is provided in Koopmann
\etal ~(2001).

As in previous papers in the series, no correction was made for 
contamination by the 2 [NII] lines ($\lambda$ $\lambda$=6548.1, 6583.8 \AA) 
that also lie within the H$\alpha$ filter bandpass. 
The ratio of [NII] to H$\alpha$ flux in HII regions has been estimated
by Kennicutt (1992) to be 0.53 in the median. It appears to be dependent upon
galaxy luminosity (e.g., Jansen \etal~ 2000) such that fainter galaxies
have a smaller [NII] contamination.
See also James \etal ~(2005) for a complete literature review.
James \etal ~(2005) probe the radial
dependence of the [NII]/H$\alpha$ correction factor using narrow-band
H$\alpha$ and [NII] imaging of 7 spiral galaxies. They find 
lower values for the global [N II]/H$\alpha$ ratio than previous
studies and show that the value has a strong dependence on radius. The
value and the radial dependence vary widely in their small
sample. One Sc galaxy, for example, has no detectable [NII]
emission. The value and radial dependence of [NII]/H$\alpha$ are
sources of uncertainty in our fluxes and surface photometry. If
the James \etal ~galaxies are typical,
the contribution from [NII] is most significant in nuclear regions,
meaning that our profiles tend to be too bright at inner radii. 
However, without detailed knowledge of the value of N II]/H$\alpha$ as
a function of radius, we do not attempt a correction.
Hereafter we will abbreviate H$\alpha$+[NII] as H$\alpha$.

The H$\alpha$ fluxes were not corrected for the
presence of the H$\alpha$ emission line in the R continuum image, in
order to be consistent with fluxes given in Koopmann \etal ~(2001) and
Young \etal ~1996. This correction amounts to a factor of 4-5\% depending
on the H$\alpha$ filter used.

These isolated sample galaxies are located at many different galactic
latitudes and it was necessary to apply a correction for galactic
extinction. Absorption values in the B filter listed in the RC3 were
scaled to R and H$\alpha$ using the values given in Rieke \& Lebofsky
(1985).  A${_R}$ values are listed in Column (9) of Table~\ref{tabsample}.
\section{Atlas of Isolated Galaxies}
\label{atlas}

R and H$\alpha$ images and surface photometry of the 29 isolated
galaxies are presented in Figure~\ref{isoim}.  Galaxies are ordered by
right ascension, with morphological information and size scales
indicated on the plots. In the lefthand column, the images are
displayed on a log scale, with north up and east to the left. In the
righthand column, R (solid) and H$\alpha$ (dotted) surface photometry
profiles are plotted as a function of radius in arc seconds.  The
H$\alpha$ profiles are displaced from the R profiles using an
arbitrary zeropoint of 18.945. The H$\alpha$ profile is plotted out to
the radius containing the outermost resolved HII region.  Radial
parameters derived from the surface photometry (as described in
Section~\ref{sp}) are indicated in the plots, as explained in the
figure caption.

\begin{figure}
\caption
{The R and H$\alpha$ images and surface photometry.  Galaxies are
ordered according to right ascension.  The images are displayed on a
log scale, with north up and east to the left. The tickmarks on the
images are spaced by 1.0 arcmin, and the solid line on each image
represents 1 arcmin, with the corresponding scale in kpc given for
each galaxy. Sky around some of the larger field-of-view images was
cropped; refer to Tables 3 and 4 for the actual size of the image
frames.  The RSA/CA morphology class is noted in the righthand corner
of the R image and the RSA/CA and RC3 morphological types are
indicated in the surface photometry plots.  In the surface photometry
plots, the R (solid) and H$\alpha$ (dotted) profiles are plotted as a
function of radius in arc seconds.  The H$\alpha$ profiles were
superposed using an arbitrary zeropoint of 18.945. A solid line
indicating 1 kpc is provided below the morphological types.  The
isophotal radius at 24 mag arcsec$^{-2}$, r$_{24}$ and the disk
scalelength, r$_d$, are indicated with arrows. An error bar for the R
profile is given at r$_{24}$, and the R profile ends where the noise
becomes greater than the signal. The H$\alpha$ profile is cut at the
radius of the outermost HII region. Diamonds on the H$\alpha$ profile
indicate annuli for which the sky uncertainty was greater than the
azimuthally averaged signal. The solid dot on the H$\alpha$ profile is
plotted at the 17 x 10$^{-18}$ isophotal radius. \label{isoim}}
\end{figure}

\subsection{Surface Photometry}
\label{sp}
Surface brightness profiles were obtained from the images using a surface
photometry program written in the IDL programming language.   
The full procedure is described in Koopmann \etal ~(2001); this section
presents a brief summary of the approach.

The center of the galaxy was derived using centroiding via the \it
center \rm task in IRAF. The axial ratio and position angle were
derived from fitting to the outer isophotes.  The center, axial ratio,
and position angle were held fixed in the determination of the
profile.

The R radial profiles were calculated until the error in the sky
background exceeded the signal.  The H$\alpha$ surface photometry was
halted just outside the radius of the outermost HII region. For
galaxies with faint outer disk star formation or galaxies with
scattered background light or poor flat fields, this radius was beyond
the point in the profile where sky uncertainty errors overwhelmed the
azimuthally averaged signal.  We indicate this radius in the plots of
radial profiles given in Figure~\ref{isoim}, but plot the H$\alpha$
radial profile to the outermost HII region to emphasize the extent of
star formation.

All profiles were corrected to face on assuming complete transparency in the
disk (see discussion in Koopmann \etal ~2001). 
Profiles of individual galaxies are plotted in the right-hand columns of 
Figure~\ref{isoim}. The H$\alpha$ profiles were arbitrarily
scaled to the R profile using a zeropoint of 18.945.
The morphological type of each galaxy is indicated. 

The main contributor of noise in the radial profile is the uncertainty in the 
sky level. There are several other systematic errors which have not been
included in the error bars. The error in the flux calibration contributes 
$\sim$ 5\%. In H$\alpha$, systematic errors due to the continuum subraction
and contribution of [N II] are typically 20-30\%. See Figure 7 of 
Koopmann \etal ~(2001) for a graphical example of the uncertainty in the
H$\alpha$ surface photometry.

Table~\ref{isorpara} presents quantities derived from the R surface
photometry. The columns provide: (1) Name of galaxy, (2) axial ratio
and, in parentheses, inclination calculated using the Hubble (1926)
conversion (see Koopmann \etal ~2001), (3) position angle, (4)
r$_{24}$, the radius in units of arcsecs at the 24 R mag/arcsec$^2$
isophote, (5) the magnitude within r$_{24}$, (6) r$_{25}$, the radius
in units of arcsecs at the 25 R mag/arcsec$^2$ isophote, (7) the
magnitude within r$_{25}$, (8) the central R light concentration
parameter, (9) the disk scalelength, and (10) the uncertainty in the
disk scalelength.including axial ratios, position angles, isophotal
radii and magnitudes, and disk scalelengths.  The central R light
concentration parameter given in Column (8) is defined analogously to
Abraham \etal ~(1994), as
$$\rm C30=\frac{F_R(0.3r_{24})}{F_R(r_{24})}$$ where F$_R$(r$_{24}$)
is the flux in R measured within the r$_{24}$ isophote and
F$_R$(0.3r$_{24}$) is the flux within the 0.3r$_{24}$ isophote.

\begin{figure}
\includegraphics[scale=0.5]{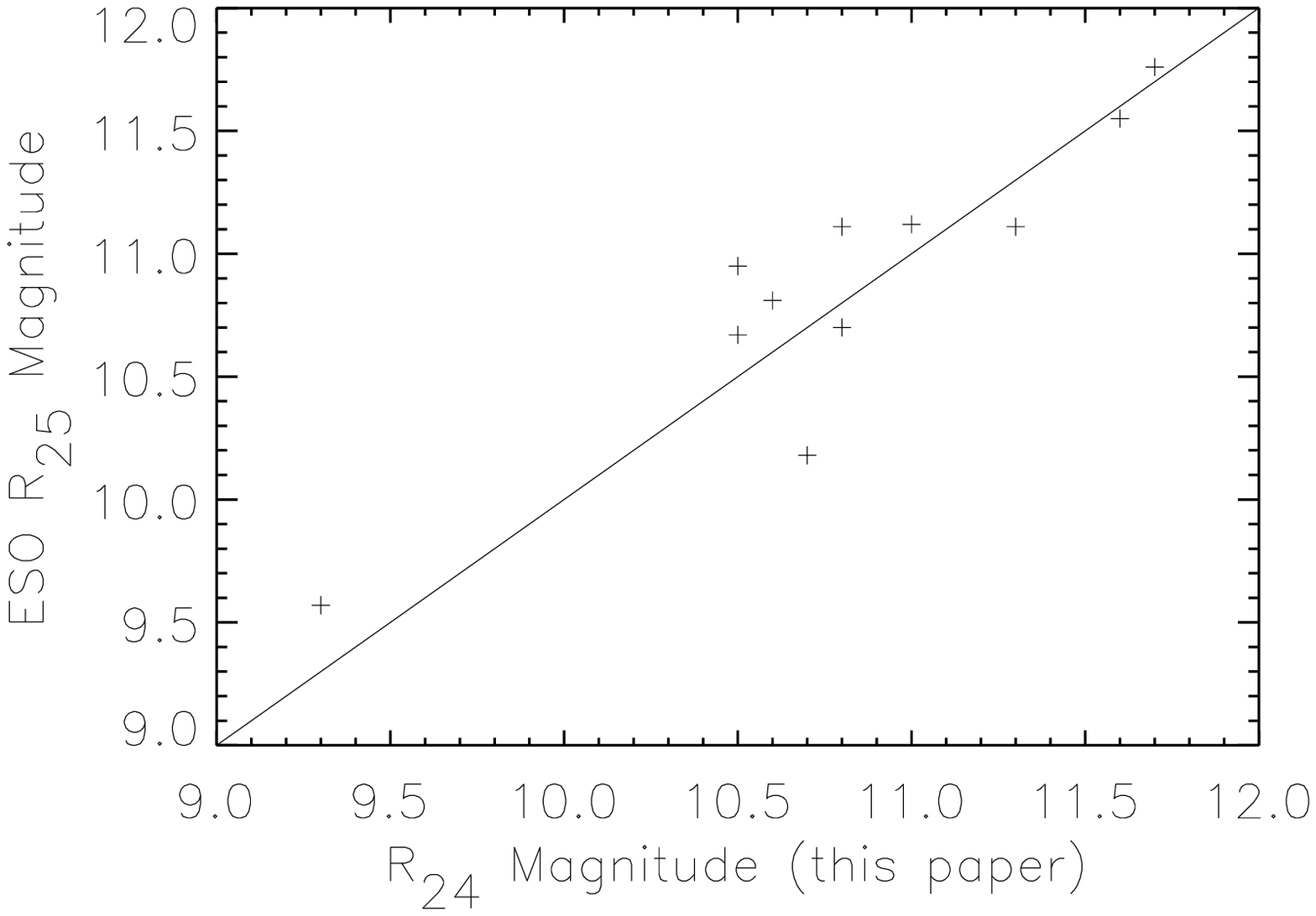}
\includegraphics[scale=0.5]{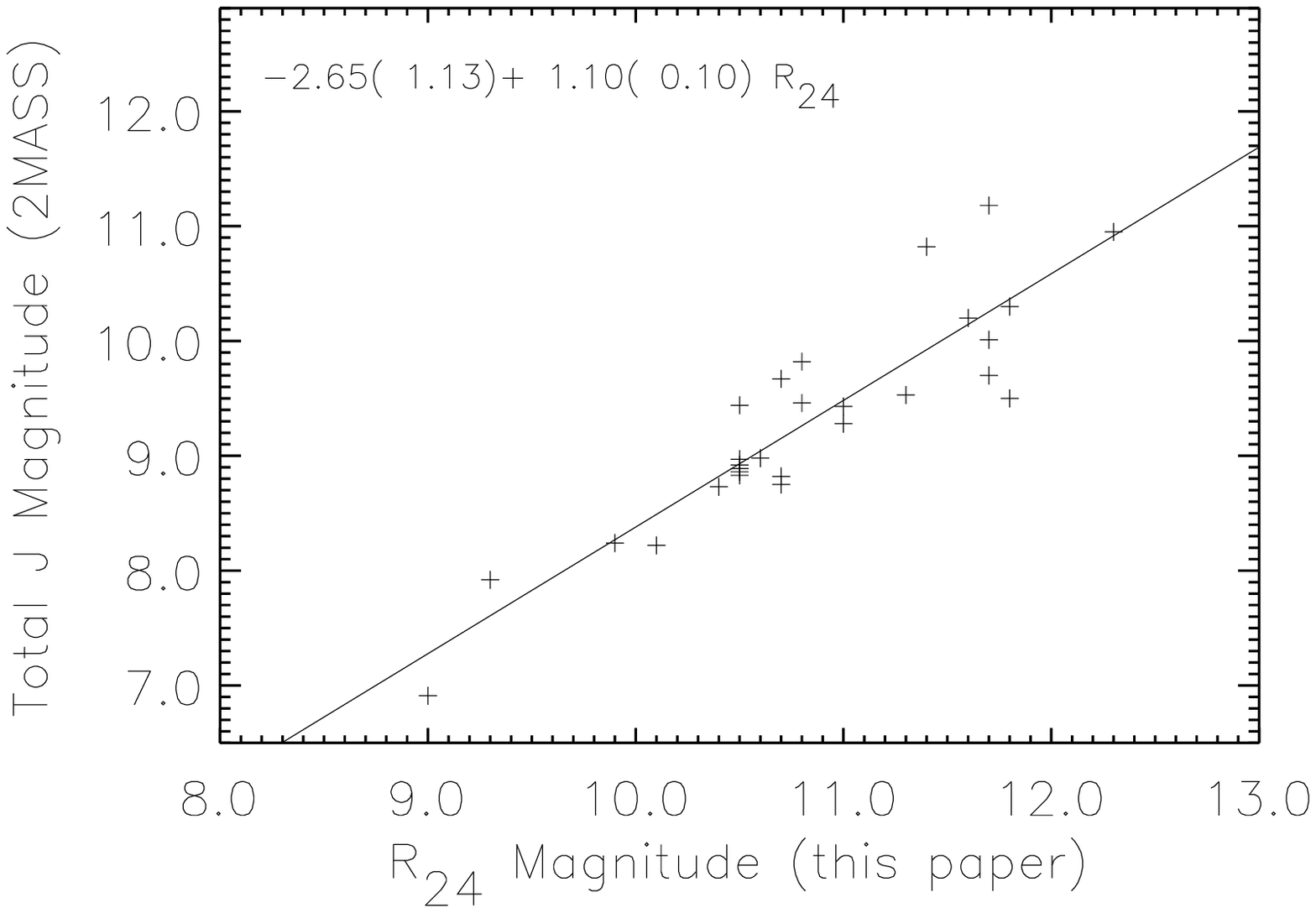}
\caption
{The R photometry given in this paper is correlated with other sources
of photometry.
(a) Comparison of isophotal R magnitudes given in this paper and in  
Lauberts \& Valentijn (1989).
for 11 galaxies in common. The solid line shows a 
one-to-one correlation. The scatter in a best-fit line is 15\%.
(b) Comparison between our isophotal R magnitudes and 2MASS total J magnitudes.
The solid line shows a linear least-squares fit to the data. The equation of
the line is given in the plot. The scatter in the fit is 10\%.  
\label{photcomp}}
\end{figure}

Table~\ref{hapar} presents quantities derived from the H$\alpha$ surface
photometry. The information listed in the columns is: (1) Name of galaxy, 
(2) log of the total H$\alpha$ flux in units of 
10$^{-18}$ erg cm$^{-2}$ s$^{-1}$, with uncertainty based on the 
background sky error combined in quadrature with an estimated 3\$ error
in the continuum subtraction factor,
(3) H$\alpha$ equivalent width in \AA, as defined below, with uncertainty,
(4) semimajor radius in arcseconds of the outermost HII region,
(5) radius in arcseconds which contains 95\% of the total H$\alpha$ flux, 
(6) radius in arcseconds at which the surface brightness falls to 
17 x 10$^{-18}$ erg cm$^{-2}$ s$^{-1}$ arcsec$^{-1}$,
(7) log of the H$\alpha$ flux within r$_{H\alpha17}$ 
in units of 1 x 10$^{-18}$ erg cm$^{-2}$ s$^{-1}$, and
(8) the H$\alpha$ concentration, as defined below.

The H$\alpha$ equivalent width given in Column (3) 
was derived from the H$\alpha$ and R photometry via the formula

\begin{displaymath}
EW = \frac{F_{H\alpha}}{kF_R} \delta \lambda,
\end{displaymath}

\noindent
where $F_R$ is the R flux, $k$ is the scaling factor used in the continuum 
subtraction (typically 0.04-0.05), 
and $\delta \lambda$ is the width of the H$\alpha$ filter in \AA.
A linear fit to the H$\alpha$ equivalent width and the NMSFR parameter
produces:
$$ EW = 1476 (\pm 83) \left (\frac{F_{H\alpha}}{F_{R}(r_{24})} \right )$$

Column (8) of Table~\ref{hapar} provides
the H$\alpha$ light concentration parameter, defined as 

\begin{displaymath}
CH\alpha = \frac{F_{H\alpha}(0.3r_{24})}{F_{H\alpha}},
\end{displaymath}

\noindent where F$_{H\alpha}$ is the total H$\alpha$ flux 
and F$_{H\alpha}$(0.3r$_{24}$) is the flux within the 0.3r$_{24}$ isophote.
A CH$\alpha$ of 1 indicates that all of the H$\alpha$ emission is
located within 0.3r$_{24}$.

\section{Comparisons to Other Sources of Photometry}

\subsection{Broadband R}

Few previous photometric measurements at red wavelengths exist for this
set of spirals. Lauberts \& Valentijn (1989) contains 
isophotal magnitudes for 11 galaxies in common. These magnitudes are
correlated with our magnitudes, as shown in Figure~\ref{photcomp}a. The solid
line indicates a one-to-one matching. The scatter in a least-squares fit 
is 15\%. 

20 of our galaxies have V magnitudes listed in the RC3, and these 
magnitudes are correlated with our isophotal R magnitudes
with a scatter of about 12\%.
Eleven of our sample galaxies have been observed in I-band by 
Haynes \etal ~(private communication). We find a 12\% scatter between the
R and I photometry.
All the sample galaxies have been observed by 2MASS. Total and
isophotal J magnitudes were extracted from the 2MASS archive using the
Large Galaxy Atlas (Jarrett \etal ~2003) where
possible. Figure~\ref{photcomp}b shows the correlation between our R
magnitudes and the total J magnitude from 2MASS. The scatter in the
best-fit line is 10\%.  (A similar scatter is seen in comparisons of
the ESO catalog R and 2MASS J photometry.)  The correlation is similar
to that obtained using our Virgo Cluster sample galaxies (Koopmann
\etal ~2001), although the Virgo Cluster spirals show somewhat less
scatter (7\%).  Results are similar when isophotal J magnitudes are
used in place of the total J magnitudes.

While the scatter in the least-square fits are similar for all 4 
comparison bands, we see few galaxies with consistently bright or faint
values. Three galaxies show trends in
at least 3 bands: our photometry of IC 5273 falls 
consistently below the trend line compared to other bands, while
NGC 3705 and NGC 986 fall consistently above.

\subsection{H$\alpha$}
\begin{figure}
\includegraphics[scale=0.5]{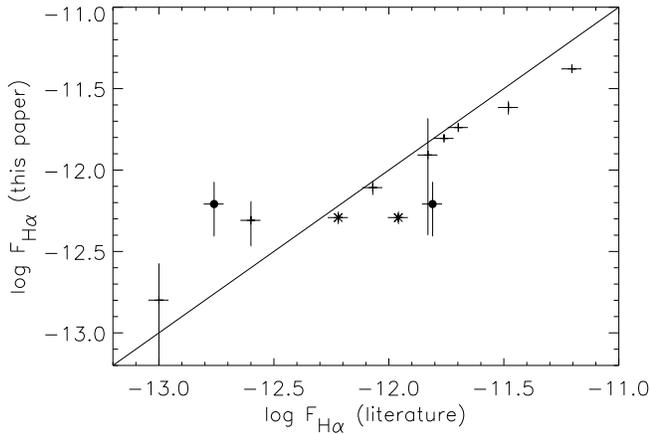}
\caption
{The H$\alpha$ photometry given in this paper compared with literature
sources as compiled by Kennicutt \& Akiyama (in prep). Our H$\alpha$
fluxes were placed on the same scale as Kennicutt \& Akiyama, by removing
the galactic extinction correction factor and adding an estimated 5\% 
correction factor to account for the presence of the H$\alpha$ line
emission in the R filter. 
The solid line shows a one-to-one correlation. Error bars on the
H$\alpha$ photometry in this paper are derived from the sky error 
combined in quadrature with an estimate of a 3\% error in the
continuum subtraction factor. Even a 3\% error in the continuum subtraction 
factor can result in a large error in the H$\alpha$ flux for galaxies
with relatively weak H$\alpha$ emission, as seen for two galaxies 
(from left to right, NGC 5273 and IC 356). Kennicutt \& Akiyama estimate
a typical error in literature values of 5-15\%; the error bars shown in the 
plot are at the 10\% level.
Cross symbols show those galaxies with one literature measurement.
Differences between authors can be much larger than estimated errors,
as seen for two galaxies with two measurements in the literature:
NGC 7098 (solid circles) and UGC 3580 (asterisks).
\label{haphotcomp}}
\end{figure}

H$\alpha$ fluxes have been measured previously for 10 galaxies,
as compiled by Kennicutt \& Akiyama (in prep.) and generously
provided to us in advance of publication. Kennicutt \& Akiyama have
placed the H$\alpha$ fluxes of approximately 2600 galaxies in the
literature on the same scale, correcting for systematic factors such as 
broad- vs narrow-band continuum filters and [N II]
contamination, and finding average fluxes for galaxies with multiple 
measurements. The comparison between the scaled
literature fluxes and our fluxes is shown in Figure~\ref{haphotcomp}.
References for the literature values can be found in Kennicutt \& Akiyama.
To place our fluxes on the same scale as Kennicutt \& Akiyama, 
the galactic extinction correction factor was removed and a 5\%
correction factor was added to account for the presence of the H$\alpha$ 
line emission in the R filter. In the plot, galaxies with one literature 
measurement are represented by cross symbols, while galaxies with two
literature measurements are represented by solid circles (NGC 7098) and
asterisks (UGC 3580). There is general agreement in the fluxes, although
our fluxes may be systematically smaller. On the
other hand, H$\alpha$ fluxes from different sources
can differ by far more than the internal
errors, as seen for the two galaxies with two literature measurements.

\section{Summary}

Broadband--R and H$\alpha$ images, surface photometry, and
quantitative parameters describing central concentrations and
star formation rates have been
obtained for a sample of 29 isolated S0-Scd spiral galaxies. Further
analyses of the spacial distribution of star formation in these galaxies
are provided in several papers.
The properties of these galaxies are quantitatively 
compared to those of Virgo Cluster
galaxies in Koopmann \& Kenney (1998; 2004a; 2004b).
H$\alpha$ scale lengths are presented and compared to Virgo Cluster and
other field galaxies in Koopmann, Haynes, \& Catinella (2005).

\acknowledgements
The funding for the research on the Virgo cluster
and isolated spiral galaxies was provided by NSF grant AST-9322779.
We thank R.B. Tully for providing us with the online version of the Nearby
Galaxies Catalog and software to access it, P. Fouqu\'{e} for providing us 
with a list of isolated galaxies, and R. Pogge for excellent advice on 
obtaining  the best continuum subtractions. 
We are grateful to Sydney Barnes, Shardha Jogee, Jerry Orosz, Sally Cersosimo,
and Charles Bailyn for obtaining many of the images in this paper during
Yale service observing runs. This work was substantially aided by
observing support from the Kitt Peak and Cerro Tololo staffs.
Martha Haynes, Christopher Springob, Karen Masters, and the Cornell 
Extragalactic Group are gratefully acknowledged for their aid in 
derivation of updated distances and HI deficiencies.
This publication makes use of data products from the Two Micron All Sky Survey,
which is a joint project of the University of Massachusetts and the Infrared 
Processing and Analysis Center/California Institute of Technology, funded by 
the National Aeronautics and Space Administration and the National Science 
Foundation.
This research has made use of the NASA/IPAC Extragalactic Database (NED)
which is operated by the Jet Propulsion Laboratory, California Institute
of Technology, under contract with the National Aeronautics and Space
Administration. The authors and maintainers of the IRAF and STSDAS software
packages and instruction manuals are also gratefully acknowledged.

\clearpage
\begin{landscape}

\begin{deluxetable}{llrcclrcrrcl}
\tabletypesize{\scriptsize}
\tablecaption{Properties of Observed Isolated Galaxies}
\tablewidth{0pt}
\tablehead{
\colhead{Name}&
\colhead{RA (2000)}&
\colhead{Dec (2000)}&
\colhead{RSA/CA}&
\colhead{RC3}&
\colhead{B$_T^O$}&
\colhead{v$_{he}$}&
\colhead{Dist}&
\colhead{Ext}&
\colhead{HI}&
\colhead{$\rho_{NBG}$}&
\colhead{NBG}\\
\colhead{}&
\colhead{(h m s)}&
\colhead{(d m s)}&
\colhead{}&
\colhead{}&
\colhead{}&
\colhead{(km s$^{-1}$)}&
\colhead{(Mpc)}&
\colhead{(A$_B$)}&
\colhead{Def}&
\colhead{(gal Mpc$^{-3}$)}&
\colhead{Affil}\\
\colhead{(1)}& \colhead{(2)}&\colhead{(3)} &\colhead{(4)} &\colhead{(5)} &\colhead{(6)} & \colhead{(7)} & \colhead{(8)} & \colhead{(9)} &\colhead{(10)} &\colhead{(11)} &\colhead{(12)}
}
\startdata
NGC 578  & 01 30 29.1 &-22 40 03  & Sc(s)II     &SAB(rs)c            &11.17 & 1630& 26.2 &.02&.17&0.07 & at-large \\
NGC 613\tablenotemark{a}& 01 34 17.4 &-29 24 58  & SBb(rs)II&SB(rs)bc&10.53 & 1475& 23.9 &.02 &.40& 0.07& at-large\\
NGC 986  &  02 33 34.2 &-39 02 40 & SBb(rs)I-II&SB(rs)ab             &11.45 & 2005& 32.2 &.03&.58& 0.06 & at-large\\
NGC 1249 &  03 10 06.4 &-53 20 17  &SBc          &SB(s)cd            &11.64 & 1072& 17.4 &0&.03& 0.27 & at-large\\
 IC  356 &  04 07 46.5 &+69 48 45 &-&SA(s)ab pec                     &10.17 &  895& 14.6 &1.2&-.17 & 0.08 & association\\
NGC 2090 &  05 47 02.3 &-34 15 05 &Sc(s)II      &SA(rs)c             &11.45 &  931& 15.6 &0&-.15 &0.14 & association \\ 
NGC 2196 &  06 12 09.5 &-21 48 23 &Sab(s)I      &SA(s)ab         &11.38 & 2321& 26.8 &.45&.31&0.10& association  \\
UGC 3580\tablenotemark{a}& 06 55 31.0 &+69 33 49&-&SA(s)a pec:       &12.20 & 1201& 18.7 &.19&.10&0.08 & at-large\\
NGC 2525\tablenotemark{a}& 08 05 37.9 &-11 25 40  &SBc(s)II&SB(s)c   &11.55 & 1581& 25.5 &.40&.45&0.06 & at-large\\
NGC 2712 &  08 59 30.6 &+44 54 51  &SBb(s)I-II   &SB(r)b:            &12.19 & 1815& 26.4 &.04&.14 & 0.17 & at-large\\
NGC 2787\tablenotemark{a} & 9 19 18.5 & +69 12 12 &SB0/a&SB0(r)+     &11.61 &  696& 12.0 &.17&0.50 & 0.06& at-large\\ 
NGC 2950 &  9 42 35.1 & +58 51 05  &RSB0$_{\frac{2}{3}}$&(R)SB(r)0   &11.80 & 1337& 20.1 &.03&$>$1 &0.22 & at-large\\ 
NGC 3329 & 10 44 39.5 &+76 48 34  &Sab         &(R)SA(r)b:          &12.57 & 1812& 27.2 &.03&- & 0.20& association \\
NGC 3359 & 10 46 36.5 &+63 13 24 &SBc(s)I.8 pec&SB(rs)c              &10.68 & 1013& 16.0 &0&-.13& 0.21 & at-large\\
NGC 3414 & 10 51 16.2 & +27 58 30  &S0$_{\frac{1}{2}}$(0)/a&S0pec    &11.86 & 1414& 21.4 &0&1.1 & 0.28& association \\
NGC 3673 & 11 25 12.7 &-26 44 11  &SBb(rs)I-II& SB(rs)b               &11.81 & 1938& 31.6 &.27&.46&.15& association \\
NGC 3705 & 11 30 07.6 &+09 16 36 &Sb(r)I-II    &SAB(r)ab             &11.25 & 1017& 16.5 &.12&.67& 0.27 &association \\
NGC 3887\tablenotemark{a}& 11 47 04.8 &-16 51 16&SBbc(s)II-III&SB(r)bc&11.19& 1209& 20.2 &.05&.03&0.08 & at-large\\
NGC 3941 & 11 52 55.3 & +36 59 11  &SB0$_{\frac{1}{2}}$/a&SB(s)0     &11.25 &  928& 14.6 &0&.40& 0.29& association \\ 
NGC 4597 & 12 40 12.8 &-05 47 59 &SBc(r)III:   &SB(rs)m              &12.21 & 1043& 17.4 &.03&-.08&0.19& at-large\\
NGC 4800 & 12 54 37.7 &+46 31 51 &Sb(rs)II-III &SA(rs)b              &12.13 &  891& 14.1 &0&.28 & 0.17& at-large\\
NGC 4984 & 13 08 57.2 &-15 30 59 &Sa(s)        &(R)SAB(rs)0+         &12.03 & 1206& 20.1 &.07&1.1 & 0.26 & at-large\\ 
NGC 5273 & 13 42 08.3 & +35 39 15  &S0/a         &SA(s)0             &12.38 & 1054& 16.3 &0&$>$1& 0.15& association \\
NGC 5334\tablenotemark{a} & 13 52 54.4 &-01 06 52&SBc(rs)II&SB(rs)c: &11.62 & 1383& 22.2 &0.15&.40&0.16& at-large \\
NGC 5669 & 14 32 43.8 &+09 53 29  &SBc(s)II     &SAB(rs)cd           &11.79 & 1371& 21.6 &.02&.17&0.29& at-large \\
NGC 7098 & 21 44 15.4 &-75 06 44 &-            &(R)SAB(rs)a           &11.63 & 2357& 37.5 &.40&.29 & 0.07& at-large \\
NGC 7141 & 21 52 14.1 &-55 34 10  &-            &SAB(rs)bc      &12.21 & 2978& 47.7 &0&-&0.18 & at-large \\
 IC 5240 & 22 41 52.3 &-44 46 04  & SBa(r)      &SB(r)a              &12.29 & 1777& 29.8 &0&.48&0.18 & at-large \\
 IC 5273 & 22 59 26.5 &-37 42 12 &SBc(s)II-III &SB(rs)cd:            &11.55 & 1301& 22.3 &.04&.08 & 0.21& at-large \\
\enddata
\label{tabsample}
\tablenotetext{a}{Galaxy also in Fouqu\'{e} list}
\end{deluxetable}
\clearpage
\end{landscape}

\begin{deluxetable}{lllccrrcccrrrr}
\tabletypesize{\scriptsize}
\tablecaption{Isolated Galaxies Observing Log}
\tablewidth{0pt}
\tablehead{
&&&\multicolumn{2}{c}{Filter}&\multicolumn{2}{c}{Exposure}&\multicolumn{2}{c}{X}&FWHM&\multicolumn{2}{c}{$\sigma_b$}&\multicolumn{2}{c}{$\delta s$}\\
\colhead{Name}&
\colhead{Date}&
\colhead{Telescope/Chip}&
\colhead{R}&
\colhead{H$\alpha$}&
\colhead{R}&
\colhead{H$\alpha$}&
\colhead{R}&
\colhead{H$\alpha$}&
\colhead{arcmin}&
\colhead{R}&
\colhead{H$\alpha$}&
\colhead{R}&
\colhead{H$\alpha$}\\
\colhead{(1)}& \colhead{(2)}&\colhead{(3)} &\colhead{(4)} &\colhead{(5)} &\colhead{(6)} & \colhead{(7)} & \colhead{(8)} & \colhead{(9)}&\colhead{(10)}&\colhead{(11)}&\colhead{(12)}&\colhead{(13)}&\colhead{(14)}
}
\startdata
NGC 578 & 1994 Aug 18& CT9/TEK2K& R&H$\alpha$6& 2 x 450 & 3 x 2400& 1.01&1.01 &1.8&530&63&188&2\\
NGC 613 & 1992 Aug 18& CT9/TEK1K& R&\dots     &     300 &\dots    & 1.11&\dots&1.7&1420&\dots&438&\dots\\
        & 1992 Aug 17& CT9/TEK1K& \dots&H$\alpha$6&\dots& 3 x 2400&\dots & 1.04&1.7&\dots&138&\dots&4\\  
NGC 986 & 1994 Aug 14& CT9/TEK2K& R&H$\alpha$6&     450 & 3 x 2400& 1.11& 1.05& 1.5& 313& 50& 188&4\\
NGC 1249& 1992 Aug 20& CT9/TEK1K& R&H$\alpha$6&     600 & 5 x 1200& 1.17& 1.14& 1.1& 640& 56& 188&4\\
IC 356  & 1995 Feb 06& KP9/t2ka & R&H$\alpha$3& 3 x 300 & 3 x 1500& 1.27& 1.29& 1.4& 410& 43& 432&2\\
NGC 2090& 1993 Mar 01& CT9/TEK2K& R&H$\alpha$5&     900 & 3 x 2400& 1.33& 1.05& 1.6& 525& 30& 188&4\\
NGC 2196& 1995 Mar 03&CT15/TEK1K& R&H$\alpha$6& 4 x  90 & 4 x  600& 1.30& 1.17& 1.6& 380& 27& 135&3\\
UGC 3580& 1995 Feb 02& KP9/t2ka & R&H$\alpha$3& 2 x 450 & 3 x 1800& 1.29& 1.27& 1.4& 162& 50&  65&2\\
NGC 2525& 1994 Feb 24& CT9/TEK2K& R&H$\alpha$6& 3 x 450 & 3 x 2400& 1.07& 1.06& 1.6& 593&143& 310&4\\
NGC 2712& 1995 Feb 03& KP9/t2ka & R&H$\alpha$4& 2 x 450 & 3 x 1200& 1.15& 1.10& 1.5& 164& 50&  65&3\\
NGC 2787& 1995 Feb 03& KP9/t2ka & R&H$\alpha$3& 4 x 150 & 3 x 1500& 1.66& 1.56& 1.6& 302& 26& 108&3\\
NGC 2950& 1995 Feb 04& KP9/t2ka & R&H$\alpha$3& 5 x  72 & 3 x  900& 1.29& 1.24& 1.6& 260& 30&  65&4\\
NGC 3329& 1995 Feb 02& KP9/t2ka & R&H$\alpha$4& 2 x 300 & 3 x 1800& 1.51& 1.47& 1.5& 260& 21&  65&2\\
NGC 3359& 1995 Feb 05& KP9/t2ka & R&H$\alpha$3& 2 x 450 & 3 x 1200& 1.51& 1.43& 1.4& 310& 10&  65&2\\
NGC 3414& 1995 Feb 06& KP9/t2ka & R&H$\alpha$4& 5 x 120 & 3 x  900& 1.37& 1.25& 1.6& 260& 20& 108&3\\
NGC 3673& 1995 Mar 29&CT15/TEK1K& R&H$\alpha$6& 4 x  90 & 4 x  600& 1.11& 1.09& 1.5& 350& 27&  81&3\\
NGC 3705& 1995 Feb 03& KP9/t2ka & R&H$\alpha$3& 4 x  70 & 4 x  900& 1.44& 1.44& 1.4& 390& 32&  43&1\\
NGC 3887& 1994 Feb 25& CT9/TEK2K& R&H$\alpha$6& 2 x 450 & 3 x 2400& 1.16& 1.09& 2.6& 535& 90& 128&4\\
NGC 3941& 1995 Feb 04& KP9/t2ka & R&H$\alpha$3& 5 x  75 & 4 x  750& 1.43& 1.32& 1.4& 280& 26& 130&4\\
NGC 4597& 1993 Mar 02& CT9/TEK1K& R&H$\alpha$6&     900 & 3 x 2400& 1.18& 1.13& 1.7& 421& 45&  64&2\\
NGC 4800& 1995 Feb 06& KP9/t2ka & R&H$\alpha$3& 3 x 240 & 4 x  900& 1.05& 1.03& 1.2& 151& 22&  65&2\\
NGC 4984& 1993 Mar 01& CT9/TEK2K& R&H$\alpha$6&     450 & 3 x 2400& 1.13& 1.04& 2.0& 287& 38& 128&4\\
NGC 5273& 1995 Feb 04& KP9/t2ka & R&H$\alpha$3& 3 x 300 & 3 x 1200& 1.01& 1.00& 1.4& 91 & 20&  43&3\\
NGC 5334& 1995 Mar 29&CT15/Tek1K& R&H$\alpha$6& 4 x  90 & 4 x  600& 1.67& 1.57& 1.1& 310& 22& 216&3\\
NGC 5669& 1995 Feb 05& KP9/t2ka & R&H$\alpha$3& 2 x 450 & 3 x 1200& 1.14& 1.11& 1.2&182 & 24&  65&2\\
NGC 7098& 1994 Aug 15& CT9/TEK2K& R&H$\alpha$6& 2 x 450 & 3 x 2400& 1.43& 1.44& 2.2& 216& 51& 255&5\\
NGC 7141& 1994 Aug 13& CT9/TEK2K& R&H$\alpha$6& 2 x 450 & 3 x 2400& 1.26& 1.35& 1.7& 264& 64& 319&5\\
IC 5240 & 1994 Aug 12& CT9/TEK2K& R&H$\alpha$6& 2 x 450 & 3 x 2400& 1.10& 1.18& 1.7& 160& 51& 160&5\\
IC 5273 & 1992 Aug 18& CT9/TEK1K& R&H$\alpha$6&     300 & 3 x 2400& 1.40& 1.07& 2.0&1500&115& 574&5\\
\enddata
\label{tabobslog}
\end{deluxetable}

\begin{deluxetable}{lcrr}
\tabletypesize{\scriptsize}
\tablecaption{Chip Codes and Characteristics}
\tablewidth{0 pt}
\tablehead{
\colhead{Detector} & \colhead{Scale} & \colhead{Size} & \colhead{FOV}\\
\colhead{} & \colhead{$^{(\prime\prime}$ pix$^{-1}$)} & \colhead{(pix)} & \colhead{$^{(\prime}$)}
}
\startdata
t2ka&0.68&2048&23.2\\
TEK2K&0.40&2048&13.7\\
TEK1K&0.40&1024&6.8\\
\enddata
\label{chip}
\end{deluxetable}

\begin{deluxetable}{llr}
\tablecaption{Filter Characteristics}
\tablewidth{0 pt}
\tablehead{
\colhead{Filter} & \colhead{$\lambda_{cent}$} & \colhead{$\delta\lambda$}\\
\colhead{} & \colhead{(\AA)} & \colhead{(\AA)}} 
\startdata
H$\alpha$3&6573&68\\
H$\alpha$4&6618&74\\
H$\alpha$5&6563&78 \\
H$\alpha$6&6606&75\\
R&6425&1540\\
\enddata
\label{filter}
\end{deluxetable}

\begin{deluxetable}{crrrrrrrrr}
\tabletypesize{\scriptsize}
\tablecaption{Parameters and Derived Quantities for the R Surface Photometry}
\tablewidth{0pt}
\tablehead{
\colhead{Name}&
\colhead{Inc}&
\colhead{PA}&
\colhead{r$_{24}$}&
\colhead{R$_{24}$}&
\colhead{r$_{25}$}&
\colhead{R$_{25}$}&
\colhead{C30}&
\colhead{r$_d$}&
\colhead{$\Delta$ r$_d$}\\
\colhead{}&
\colhead{($^{\circ}$)}&
\colhead{($^{\circ}$)}&
\colhead{($^{\prime\prime}$)}&
\colhead{(mag)}&
\colhead{($^{\prime\prime}$)}&
\colhead{(mag)}&
\colhead{}&
\colhead{($^{\prime\prime}$)}&
\colhead{($^{\prime\prime}$)}\\
\colhead{(1)}& \colhead{(2)}&\colhead{(3)} &\colhead{(4)} &\colhead{(5)} &\colhead{(6)} & \colhead{(7)} & \colhead{(8)} & \colhead{(9)} &\colhead{10)} 
}
\startdata
NGC 578 & 0.574 (57) & 110 & 117 & 10.8 & 146& 10.7 & 0.36 & 35 & 3\\
NGC 613  & 0.799 (38)  & 120 & 207 & 9.3 & 0 & 0.0 & 0.44 & 63 & 3\\
NGC 986  & 0.743 (43) & 150 & 119 & 10.7 & 144 & 10.6 & 0.49 & 38 & 3\\
NGC 1249  & 0.500 (62) & 85 & 94 & 11.6 & 121 & 11.5 & 0.32 & 40 & 3\\
 IC  356  & 0.719 (45) & 105\tablenotemark{a} & 252 & 9.0 & 0 & 0.0 & 0.52 & 81 & 10\\
NGC 2090  & 0.484 (63) & 13 & 99 & 10.5 & 173 & 10.4 & 0.50 & - & -\\ 
NGC 2196  & 0.682 (48)\tablenotemark{a}& 53\tablenotemark{a} & 102 & 10.6 & 140 & 10.5 & 0.54 & 31 & 5\\
UGC 3580  & 0.469 (64) & 5 & 60 & 12.3 & 90 & 12.1 & 0.49 & 28 & 3\\
NGC 2525  &  0.643 (51) & 75 & 100 & 10.7 & 118 & 10.7 & 0.30 & 30 & 5\\
NGC 2712  & 0.500 (62) & 0 & 64 & 11.7 & 86 & 11.7 & 0.37 & 19 & 2\\
NGC 2787  & 0.574 (57) & 130\tablenotemark{a} & 104 & 10.1 & 129 & 10.0 & 0.61 & 25 & 3\\
NGC 2950  & 0.656 (50) & 127\tablenotemark{a} & 93 & 10.5 & 113 & 10.5 & 0.73 & 26 & 2\\ 
NGC 3329  & 0.559 (58) & 133\tablenotemark{a} & 55 & 11.8 & 77 & 11.8  & 0.58 & 20 & 3\\
NGC 3359  & 0.643 (51) & 175 & 142 & 10.5 & 184 & 10.4 & 0.37 & 43 & 2\\
NGC 3414  & 0.839 (34)\tablenotemark{a} & 30 & 104 & 10.5 & 149 & 10.5 & 0.61 & - & -\\
NGC 3673  & 0.500 (62)\tablenotemark{a} & 75 & 109 & 11.0 & 139 & 11.0 & 0.41 & 33 & 3\\
NGC 3705  &  0.423 (68) & 119 & 112 & 10.7 & 146 & 10.6 & 0.35 & 32 & 2\\
NGC 3887  &  0.766 (41) & 19 & 109 & 10.5 & 130 & 10.4  & 0.40 & 27 & 3\\
NGC 3941  &  0.731 (44)\tablenotemark{a} & 10 & 98 & 9.9 & 119 & 9.9 &  0.65 & 22 & 2\\ 
NGC 4597  &  0.423 (68) & 49  & 93 & 12.0 & 120 & 11.8 & 0.21 & 40 & 6\\
NGC 4800  &  0.766 (41) & 25 & 60 & 11.0 & 78 & 11.0 & 0.56 & 19 & 5\\
NGC 4984  &  0.755 (42) & 15\tablenotemark{a} & 115 & 10.4 & 159 & 10.3 & 0.65 & 40 & 4\\ 
NGC 5273  &  0.866 (31) & 10 & 68 & 11.8 & 84 & 11.7 & 0.42 & 20 & 3\\
NGC 5334  & 0.719 (45)  & 15 & 101 & 11.4 & 126 & 11.3 & 0.27 & 40 & 4\\
NGC 5669  & 0.707 (46) & 50 & 90 & 11.7 & 117 & 11.5 & 0.34 & 31 & 4\\
NGC 7098 & 0.574 (57) & 70 &  135 & 10.5 & 162 & 10.5 & 0.57 & 42 & 4\\
NGC 7141  & 0.707 (46) & 0\tablenotemark{a} & 77 & 11.7 & 124 & 11.5 & 0.39 & 32 & 4\\
 IC 5240  & 0.643 (51) & 100 & 80  & 11.3 & 99 & 11.3 & 0.49 & 24 & 3\\
 IC 5273  & 0.602 (55) & 56  & 102 & 10.8 & 0  & 0.0  & 0.35 & 31 & 2\\
\enddata
\label{isorpara}
\tablenotetext{a}{Value different by more than 5$^\circ$ from RC3.}
\end{deluxetable}

\begin{deluxetable}{lcrrrccc}
\tabletypesize{\scriptsize}
\tablecaption{Parameters and Derived Quantities from the H$\alpha$ surface photometry}
\tablewidth{0pt}
\tablehead{
\colhead{Name}&
\colhead{log F$_{H\alpha}$} &
\colhead{H$\alpha$ EW}&
\colhead{r$_{HII}$}&
\colhead{r$_{H\alpha95}$}&
\colhead{r$_{H\alpha17}$}&
\colhead{log F$_{H\alpha17}$}&
\colhead{CH$\alpha$}\\
\colhead{}&
\colhead{}&
\colhead{(\AA)}&
\colhead{($^{\prime\prime}$)}&
\colhead{($^{\prime\prime}$)}&
\colhead{($^{\prime\prime}$)}&
\colhead{}&
\colhead{}\\
\colhead{(1)}& \colhead{(2)}&\colhead{(3)} &\colhead{(4)} &\colhead{(5)} &\colhead{(6)} & \colhead{(7)} & \colhead{(8)} 
}
\startdata
NGC 578 & -11.52 $\pm$ 0.03 & 36 $\pm$ 2 & 135 & 104 & 102 & -11.54 & 0.25 \\
NGC 613 & -11.34 $\pm$ 0.09 & 14 $\pm$ 2 & 164 & 101 & 113 & -11.34 & 0.60\\
NGC 986 & -11.63 $\pm$ 0.04 & 23 $\pm$ 2 & 108 & 82 & 91 & -11.63 & 0.49\\
NGC 1249 & -11.76 $\pm$ 0.03  & 39 $\pm$ 3 & 179 & 137 & 80 & -11.85 & 0.25\\
 IC  356 & -11.66 $\pm$ 0.17 & \ 5 $\pm$ 2 & 245 & 184 & 150 & -11.73 & 0.29 \\
NGC 2090 & -11.58 $\pm$ 0.04 & 31 $\pm$ 2 & 287 & 203 & 122 & -11.73 & 0.32\\
NGC 2196 & -11.74 $\pm$ 0.06 & 16 $\pm$ 2 & 123 & 112 & 99 & -11.80 & 0.18\\
UGC 3580 & -12.27 $\pm$ 0.04 & 25 $\pm$ 2 & 130 & 100 & 35 & -12.40 & 0.63\\
NGC 2525 & -11.49 $\pm$ 0.03 & 35 $\pm$ 3 & 88 & 69 & 79 & -11.49 & 0.20\\
NGC 2712 & -12.12 $\pm$ 0.04 & 23 $\pm$ 2 & 98 & 69 & 59 & -12.17 & 0.27\\
NGC 2787 & -13.14 $\pm$ 0.7 \ & 0.5 $\pm$ 2 & 31 & 27 & 13 & -13.31 & 1.00\\
NGC 2950 & -12.59 $\pm$ 0.5 \ & 3 $\pm$ 2 & 49 & 44 & 31 & -12.69 & 0.76\\
NGC 3329 & -12.33 $\pm$ 0.06 & 13 $\pm$ 2 & 63 & 53 & 34 & -12.42 & 0.43\\
NGC 3359 & -11.40 $\pm$ 0.02 & 12 $\pm$ 2  & 235 & 176 & 141 & -11.46 & 0.24\\
NGC 3414 & -12.89 $\pm$ 0.4 \ &\ 1 $\pm$ 2  & 33 & 28 & 19 & -13.00 & 0.99\\
NGC 3673 & -12.09 $\pm$ 0.09 & 11 $\pm$ 2 & 139 & 124 & 73 & -12.29 & 0.13\\
NGC 3705 & -11.69 $\pm$ 0.04 & 24 $\pm$ 2 & 196 & 141 & 96 & -11.77 & 0.23\\
NGC 3887 & -11.36 $\pm$ 0.03 & 39 $\pm$ 3 & 125 & 100 & 108 & -11.38 & 0.17\\
NGC 3941 & -12.50 $\pm$ 0.3 \ &\ 2$\pm$ 2  & 94 & 87 & 17 & -13.07 & 0.31\\
NGC 4597 & -11.82 $\pm$ 0.02 & 48 $\pm$ 3 & 149 & 123 & 106 & -11.87 & 0.07\\
NGC 4800 & -11.89 $\pm$ 0.04 & 20 $\pm$ 2 & 58 & 40 & 45 & -11.90 & 0.54\\
NGC 4984 & -11.99 $\pm$ 0.11 & \ 8 $\pm$ 2 &  50 & 35 & 41 & -12.00 & 0.94\\
NGC 5273 & -12.82 $\pm$ 0.4 \ & \ 3 $\pm$ 2 & 10 & 3 & 6 & -12.82 & 0.45\\
NGC 5334 & -11.82 $\pm$ 0.04 & 27 $\pm$ 2 & 131 & 115 & 102 & -11.89 & 0.16\\
NGC 5669 & -11.82 $\pm$ 0.02 & 42 $\pm$ 2 & 138 & 115 & 113 & -11.84 & 0.23\\
NGC 7098 & -12.14 $\pm$ 0.14 & \ 7$\pm$ 2  & 137 & 126 & 66 & -12.41 & 0.25\\
NGC 7141 & -12.28 $\pm$ 0.08 & 13 $\pm$ 2 & 126 & 114 & 64 & -12.43 & 0.35\\
 IC 5240 & -12.33 $\pm$ 0.14 & \ 7 $\pm$ 2 & 88 & 79 & 60 & -12.47 & 0.10\\
 IC 5273 & -11.77 $\pm$ 0.06 & \ 7 $\pm$ 2 & 89 & 60 & 70 & -11.77 & 0.38\\
\enddata
\label{hapar}
\end{deluxetable}

\end{document}